\documentclass[preprint,showpacs,preprintnumbers,amsmath,amssymb,prb]{revtex4-1}
\usepackage[version=3]{mhchem} % Formula subscripts using \ce{}
\usepackage{graphicx}% Include figure files
\usepackage{dcolumn}% Align table columns on decimal point
\usepackage{bm}% bold math
\begin{document}
\newcommand*{\mycommand}[1]{\texttt{\emph{#1}}}
\author{M.I. Katsnelson, A. Fasolino}

\affiliation {
Radboud University Nijmegen, Institute for Molecules and Materials, Heyendaalseweg 135, 6525 AJ Nijmegen, The Netherlands
         }

\title{Graphene as a prototype crystalline membrane}

%\begin{abstract}

\begin{abstract}

The understanding of the structural and thermal properties of membranes, low-dimensional flexible systems  in a space of higher dimension, is pursued in many fields from string theory to chemistry and biology.  The case of a two dimensional (2D) membrane in three dimensions is the relevant one for dealing with real materials.
 Traditionally, membranes are primarily discussed in the context of biological membranes and soft matter in general.  The complexity of these systems hindered a realistic description of their interatomic structures based on a truly microscopic approach. Therefore theories of membranes were developed mostly within phenomenological models.  
From the point of view of statistical mechanics, membranes at finite temperature are systems governed by interacting long-range fluctuations. 

Graphene, the first truly two-dimensional system consisting of just one layer of carbon atoms, provides a model system for the development of a microscopic description of membranes. In the same way that geneticists have used Drosophila as a gateway to probe more complex questions, theoretical chemists and physicists can use graphene as a simple model membrane to study both phenomenological theories and experiments.  In this Account, we review key results in the microscopic theory of structural and thermal properties of graphene and compare them with the predictions of phenomenological theories.  The two approaches are in good agreement for the various scaling properties of correlation functions of atomic displacements.  However, some other properties, such as the temperature dependence of the bending rigidity, cannot be understood based on phenomenological approaches.  We also consider graphene at very high temperature and compare the results with existing models for two-dimensional melting.  The melting of graphene presents a different scenario, and we describe that process as the decomposition of the graphene layer into entangled carbon chains. 
\end{abstract}

%\end{abstract}
\maketitle
\includegraphics[width=1\linewidth,angle=0,clip]{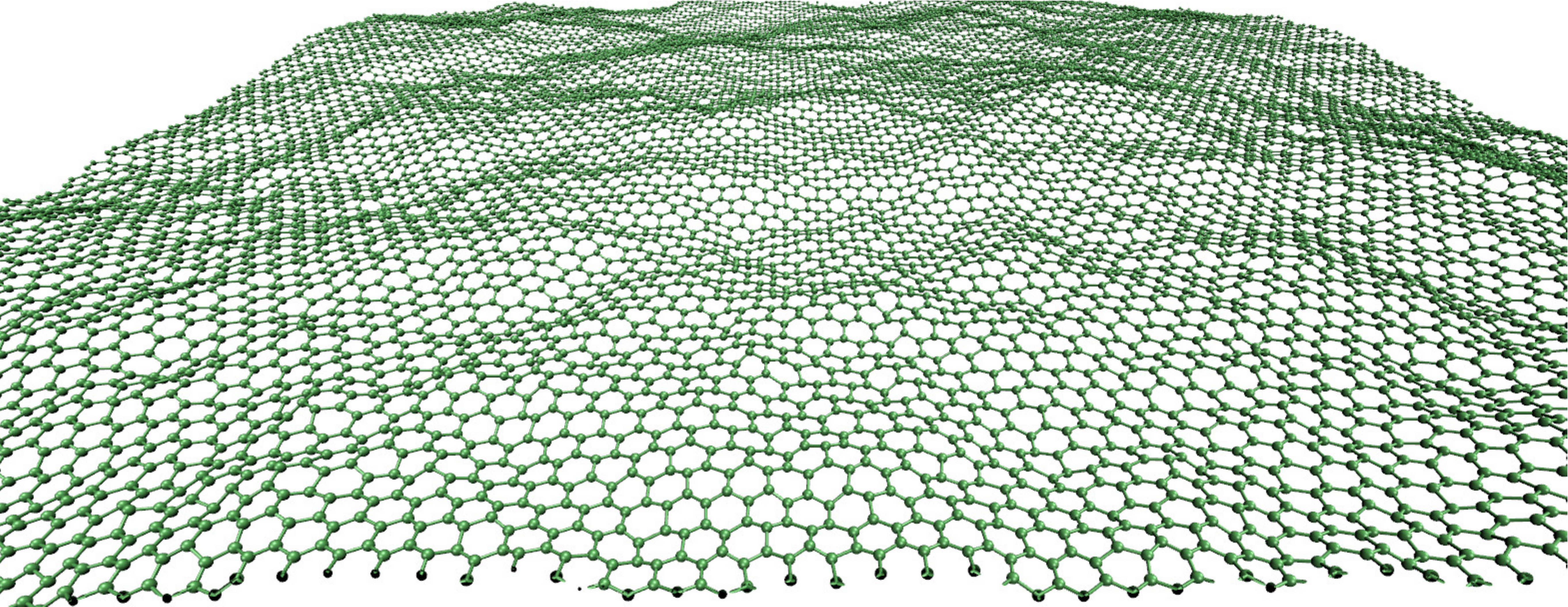}
\section{Introduction}
Understanding the structural and thermal properties of two dimensional (2D) systems is of 
great interest in many fields including mechanics, statistical  
physics, chemistry and biology.~\cite{nelsonbook}. 
Traditionally, it was discussed mainly in the context of biological membranes and soft condensed matter. The complexity of these systems hindered any truly microscopic approach based on a realistic description of interatomic interactions. Phenomenological theories of membranes\cite{nelsonbook,bookKats} based on elasticity\cite{LL}  reveal nontrivial scaling behavior of physical properties, like in- and out-of-plane atomic displacements.  In three-dimensional (3D) systems, this type of behavior takes place only close to critical points\cite{Ma}, whereas in 2D this occurs at any finite temperature. 
The discovery of graphene~\cite{Geim}, the first truly 2D crystal made of just one layer of carbon atoms, provides a model system 
for which an atomistic description becomes possible. The interest for graphene has been triggered by its exceptional electronic properties (for review see~\cite{Geim,bookKats}) but the experimental observation of ripples in freely suspended graphene~\cite{Meyer} has initiated a theoretical interest also in the structural properties~\cite{our,bookKats}. Ripples or bending fluctuations have been proposed as one of the dominant scattering mechanisms that determine the electron mobility in graphene~\cite{bookKats}. Last but not least, the structural state influences the mechanical properties that  are important for numerous potential applications of graphene~\cite{KatsGeimNanolet,Makianmicromechreson,Science}.

Graphene is a crystalline membrane, with finite resistance to in plane shear deformations, contrary to liquid membranes as soap films.
Moreover, for graphene, this resistance is extremely high since the carbon-carbon bond is one of the strongest chemical bonds in nature. The Young modulus per layer of graphene is 350 N/m, an order of magnitude larger than that of steel\cite{Science,prlkostya}. Phenomenological theories just assume that the membrane thickness is negligible in comparison to the lateral dimensions. Once a microscopic treatment is allowed, one can also distinguish between, e.g., single layers and  bilayers\cite{Zak2010}.
  
The aim of this review is to summarize the contribution  of microscopic treatments of graphene as the simplest (prototype) membrane to our general understanding of 2D systems. To this purpose we first review the main results of phenomenological theories, particularly those that can be directly compared to results of atomistic approaches. 

\section{Phenomenological theory of crystalline membranes}
The standard theory of lattice dynamics is based on the harmonic approximation assuming atomic displacements  from equilibrium to be  much smaller than the interatomic distance $d$. For 3D crystals, this assumption holds up to  melting  according to the empirical Lindemann criterion.
For 2D crystals the situation is so different that Landau and Peierls suggested in the 1930's that 2D crystals cannot exist. 
Later, their qualitative arguments were made more rigorous in the context of the so-called Mermin-Wagner theorem (see references in \cite{Meyer}. Since graphene is generally considered to be a  2D {\it crystal} this point needs to be clarified first. 

\subsection{Lattice dynamics of graphene}
By aauming that atomic displacements $\vec{u}$ satisfy the condition
\begin{equation}
<\vec{u}^2_{n,j}><< d^2
\label{smallness} 
\end{equation}
where $n$ labels the elementary cell and $j$ the atoms within the elementary cell, we can expand the potential energy $V(\vec{R})$  up to  quadratic terms (harmonic approximation):
\begin{equation}
V\left(\vec{R}_{n,j}\right)= V\left(\vec{R}^{(0)}_{n,j}\right) +\frac{1}{2} 
\sum_{n,n'~i,j~\alpha\beta} 
A^{\alpha\beta}_{ni,n'j} u_{ni}^{\alpha\beta} u_{n'j}^{\alpha\beta}
\label{harmV} 
\end{equation}
where the matrix $\hat{A}$ is the force constant matrix, $\vec{R}_{nj}=\vec{R}^{(0)}_{nj}+\vec{u}_{nj}$ and $\vec{R}^{(0)}_{nj}=\vec{r}_n+ \vec{\rho}_j$ where $\vec{r}_n$ are the vectors of the 2D Bravais lattice and $\vec{\rho}_j$ are basis vectors. Lattice vibrations are then described as superposition of independent modes, called phonons, characterized by wavevector $\vec{q}$ and branch number $\xi=1,....,3\nu$ where $\nu$ is the number of atoms per unit cell. The squared phonon frequencies $\omega^2_{\xi}(\vec{q})$ are the eigenvalues of the $3\nu\times 3\nu$ dynamical matrix 
\begin{equation}
D_{ij}^{\alpha\beta}(\vec{q})=\sum_{n} \frac {A^{\alpha\beta}_{0i,nj}}{\sqrt{M_iM_j}} exp(i\vec{q}\cdot\vec{r_n})
\label{dyn_mat}
\end{equation}
where $M_j$ is the mass of atom $j$. For graphene, $M_j=M$ is the mass of the carbon atom and by symmetry $A_{i,j}^{xz}=A_{i,j}^{yz}=0$ and  $D_{1,1}^{\alpha\beta}=D_{2,2}^{\alpha\beta}$.
Translational invariance requires that no forces result from a rigid shift of the crystal, implying:
\begin{equation}
\sum_{nj} A^{\alpha\beta}_{0i,nj}=0
\label{tr_inv}
\end{equation} 
whence
\begin{equation}
D_{12}^{\alpha\beta}(\vec{q}=0)+D_{11}^{\alpha\beta}(\vec{q}=0) =0
\label{cond}
\end{equation} 

\begin{figure}
\caption{ Phonon spectrum of graphene calculated with LCBOPII. Adapted from \cite{Karssemeijer}}.
\includegraphics[width=0.5\linewidth,angle=0]{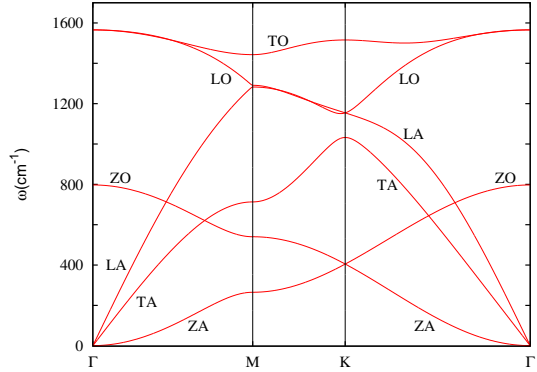}
\label{phononspectrum}
\end{figure}

Therefore, there are six phonon branches in graphene:
\begin{itemize}
\item[1)]The acoustic flexural mode ZA $(\vec{u}|| Oz)$ 
\begin{equation}
\omega_{ZA}^2(\vec{q})= D_{11}^{zz}(\vec{q})+D_{12}^{zz}(\vec{q})
\label{ZA}
\end{equation}
\item[2)] The optical flexural mode ZO  $(\vec{u}|| Oz)$
\begin{equation}
\omega^2_{ZO}(\vec{q})= D_{11}^{zz}(\vec{q})-D_{12}^{zz}(\vec{q}).
\end{equation}
\item[3),4)] Two acoustic in-plane modes, with $\omega^2(\vec{q})$ equal to the eigenvalues of 
the $2 \times 2$ matrix
\begin{equation}
D_{11}^{\alpha\beta}(\vec{q})+D_{12}^{\alpha\beta}(\vec{q})~~~ (\alpha,\beta=x,y)
\end{equation}
\item[5),6)] Two optical  in-plane modes, with  $\omega^2(\vec{q})$ equal to eigenvalues of the $2 \times 2$ matrix
\begin{equation}
D_{11}^{\alpha\beta}(\vec{q})-D_{12}^{\alpha\beta}(\vec{q})~~~ (\alpha,\beta=x,y)
\end{equation}
\end{itemize}
If the 2D wavevector $\vec{q}$ lies in symmetric directions, branches
(3)-(6) can be divided into longitudinal $\vec{e}||\vec{q}$  and transverse 
$\vec{e}\perp \vec{q}$ modes.
Due to \ref{cond}  for acoustic modes
$\omega^2 \propto q^2$ at $\vec{q} \rightarrow 0$. For the ZA mode, however, the terms in $q^2$ disappear as well and $\omega^2_{ZA}(q) \propto q^4$\cite{Lifsh52}. 
This follows from the invariance with respect to {\it rotations} of a 2D crystal as a whole in the 3D space, namely for uniform  rotations of the type
\begin{equation}
\vec{u}_{nj}=\delta\phi \vec{m} \times \vec{R}^{(0)}_{nj},
\end{equation}
where $\delta\phi$ is the rotation angle and $\vec{m}$ the rotation axis  in the $xy$-plane. These rotations should not lead to any forces or torques acting on the atoms. Hence,  
\begin{equation}
\sum_{nj} A^{zz}_{0i,nj}r_n^\alpha r_n^\beta=0~~~ (\alpha,\beta=x,y).
\label{rot_inv}
\end{equation}
It follows from \ref{rot_inv}, \ref{dyn_mat} that
\begin{equation}
\frac{\partial^2} {\partial q_\alpha \partial q_\beta} \left[D_{11}^{zz}(\vec{q})+D_{12}^{zz}(\vec{q})\right]_{\vec{q}=0} =0 
\end{equation}
and, thus, the expansion of \ref{ZA} starts with terms
of the order of $q^4$; therefore,
\begin{equation}
\omega_{ZA} \propto q^2
\label{ZA+}
\end{equation}
at $\vec{q} \rightarrow 0$. The very low frequency of $\omega_{ZA}$ for $\vec{q} \rightarrow 0$  has important consequences for the stability and thermal properties as we discuss next.
In \ref{phononspectrum} we show the phonon spectrum\cite{Karssemeijer}
calculated with the  so-called long-range carbon bond order potential (LCBOPII)\cite{Los2005} used in the atomistic simulations presented later. 

Let us consider now the case of finite temperatures. In the harmonic
approximation, the mean-square atomic displacement is
\begin{equation}
<u_{nj}^\alpha u_{nj}^\beta>=\sum_\lambda \frac{\hbar}{2N_0 M_j \omega_\lambda} \left( e_{\lambda j}^\alpha\right)^*\left( e_{\lambda j}^\beta\right) coth\left(\frac{\hbar\omega_\lambda}{2T}\right)
\label{uu}
\end{equation}
where $\lambda=(\vec{q},\xi)$ are phonon labels, $\vec{e}$ is the polarization vector and $N_0$ is the number of elementary cells.
For in-plane deformations at any finite temperature the sum in  \ref{uu} is logarithmically divergent due to the contribution of acoustic
branches with $\omega \propto q$ for $\vec{q} \rightarrow 0$. This divergence is cut at the minimal wavevector $q_{min} \sim L^{-1}$ ($L$ is the sample size), thus
\begin{equation}
<x_{nj}^2 > = <y_{nj}^2>\approx  \frac{T}{2 \pi M c_s^2} \ln \left(\frac{L}{d}\right)
\label{x2}
\end{equation}
where $c_s$ is the average sound velocity. This result  led Landau and Peierls to the conclusion that
2D crystals cannot exist. Strictly speaking, this means just the
inapplicability of the harmonic approximation, due to violation of \ref{smallness}. A  more rigorous treatment, however, does confirm this conclusion (see \cite{bookKats}). 
For $\alpha=z$, the situation is even worse, due to the much stronger divergence of ZA phonons (\ref{ZA}). One can see from \ref{uu} that
\begin{equation}
<h_{nj}^2 >   \propto \frac{T}{E_{at}}\sum_q \frac{1}{q^4} \propto \frac{T}{E_{at}} L^2
\label{h2}
\end{equation}
where $E_{at}$ is of the order of the cohesive energy. Henceforth we use the notation $h=u^z$, and denote $\vec{u}=(u_x,u_y)$ as  a 2D vector.  

\subsection{The statistical mechanics of crystalline membranes}

We have shown that the harmonic approximation cannot be applied at any finite temperature to 2D crystals neither for in-plane nor for out-of-plane modes since  the condition \ref{smallness} is violated due to divergent contributions of acoustic long-wavelengths modes with $q\rightarrow 0$. In this situation, it becomes necessary to consider anharmonic interactions between in-plane and out-of-plane modes. 
In the limit $q\rightarrow 0$, acoustic modes can be described by elasticity\cite{LL}. The corresponding effective Hamiltonian ${\cal H}$ reads
\begin{equation}
\label{eq:Free_F_1}
{\cal H} = \frac{1}{2}\int\! d^2\!x\,\left(\kappa \left(\nabla^2 h\right)^2 + \mu u_{\alpha\beta}^2+\frac{\lambda}{2}u_{\alpha\alpha}^2\right),
\end{equation}
where the deformation tensor $u_{\alpha\beta}$ is
\begin{equation}
u_{\alpha\beta} = \frac{1}{2}\left(\frac{\partial u_{\beta} }{\partial x_{\alpha}}+\frac{\partial u_{\alpha}}{\partial x_{\beta}}+\frac{\partial h}{\partial x_{\alpha}} \frac{\partial h}{\partial x_{\beta}} \right)
\label{strain}
\end{equation}
$\kappa$ is the bending rigidity and  $\mu$ and $\lambda$ are  Lam\'e coefficients.
In the deformation tensor, we have kept the nonlinear terms in $\partial h /\partial x_\alpha$ but not $\partial u_{\gamma}/\partial x_\alpha$ since out-of-plane fluctuations are stronger than in-plane ones (compare  \ref{x2} and \ref{h2}). If we neglect all nonlinear terms in the deformation tensor, then ${\cal H}$ in $\vec{q}$ representation becomes:
\begin{equation}
\label{eq:Free_F_0}
{\cal H}_0 = \frac{\kappa}{2} \sum_{\vec{q}} q^4 |h_{\vec{q}}|^2 + \frac{1}{2} \sum_{\vec{q}} \left[ \mu q^2  |\vec{u}_{\vec{q}}|^2 +\left(\lambda+\mu\right) \left(\vec{q}\cdot \vec{u}_{\vec{q}}\right)^2\right]
\end{equation}
where the subscript $0$ indicates the harmonic approximation and 
$ h_{\vec{q}}$ and $\vec{u}_{\vec{q}}$ are Fourier components of $h(\vec{r})$ and $\vec{u}(\vec{r})$, respectively, with $\vec{r}=(x,y)$.

The correlation functions in harmonic approximation are
\begin{equation}
G_0(\vec{q})= <|h_{\vec{q}}|^2>_0=\frac{T} {\kappa q^4}
\label{G0}
\end{equation}

\begin{equation}
D_0^{\alpha\beta}(\vec{q})= <u_{\alpha \vec{q}}^* u_{\beta \vec{q}}^{ }>_0= \frac{q_\alpha q_\beta}{q^2}\frac{T} {(\lambda+2\mu) q^2} +\left[ \delta_{\alpha\beta} - \frac{q_\alpha q_\beta}{q^2}\right] \frac{1} {\mu q^4}
\label{D0}
\end{equation}
where $< >_0$ means averaging with the Hamiltonian ${\cal H}_0$ (\ref{eq:Free_F_0}). 

For a surface $z=h(x,y)$ the components of the normal are
\begin{eqnarray}
 n_x&=&-\frac{\partial h}{\partial x}\frac{1} {\sqrt{1+|\nabla h|^2}}\\
n_y&=&-\frac{\partial h}{\partial y}\frac{1} {\sqrt{1+|\nabla h|^2}}\\
n_z&= &\frac{1} {\sqrt{1+|\nabla h|^2}}
\end{eqnarray}
where  $\nabla h$ is a 2D gradient.  If $|\nabla h |<<1$, the normal-normal
correlation function is related to $<|h_{\vec{q}}|^2>$ 
\begin{equation}
< \vec{n}_{\vec{q}}\vec{n}_{-\vec{q}}>= q^2  <|h_{\vec{q}}|^2> 
\label{nn}
\end{equation} 
On substituting \ref{G0} into \ref{nn} we find	
\begin{equation}
<\vec{n}_{\vec{q}}\vec{n}_{-\vec{q}} >_0 = \frac{T}{\kappa q^2}
\label{nnT}
\end{equation} 
A membrane is globally flat if the correlation function $<\vec{n}_{0}\vec{n}_{\vec{R}}>$ tends to a constant as $R\rightarrow \infty$ (normals at large distances have, on average,
the same direction).  \ref{nnT}, instead, leads to a logarithmic divergence of $<\vec{n}_0\vec{n}_{\vec{R}}>$. Moreover, the mean square in-plane and out-of-plane displacements calculated from \ref{G0} and \ref{D0} are divergent as $L\rightarrow \infty$ as already shown. 
Again, we conclude that
the statistical mechanics of 2D systems cannot be based on the harmonic approximation.
Taking into account the  coupling between  $\vec{u}$ and $h$ due to the nonlinear terms in the deformation tensor \ref{strain} drastically changes this situation. We can introduce
the renormalized bending rigidity $\kappa_R(q)$ by writing
\begin{equation}
G(\vec{q})= \frac{T} {\kappa_R(q) q^4}
\label{G}
\end{equation}
The first-order anharmonic correction to $\kappa$ is
\begin{equation}
\delta \kappa \equiv \kappa_R(q)-\kappa=\frac{3TY}{8\pi\kappa q^2}
\label{delta}
\end{equation}
where $Y=\frac{4\mu(\lambda+\mu)}{\lambda+2 \mu}$ is the 2D Young modulus\cite{nelsonbook,bookKats}.  At 
\begin{equation}
q = q^*= \sqrt{\frac{3TY} {8 \pi \kappa^2}}
\label{q*}
\end{equation}
the correction $\delta\kappa = \kappa$, and the coupling between in-plane and out-of-plane distortions cannot be considered as a perturbation.   
The value $q^*$ plays the same role as the Ginzburg criterion\cite{Ma}  in the theory of critical phenomena: below $q^*$ interactions between fluctuations dominate. Note that in the theory of {\it liquid} membranes, there is also
a divergent anharmonic correction to $\kappa$ of completely different origin\cite{nelsonbook}
\begin{equation}
\delta \kappa \approx -\frac{3T}{4\pi}\ln\left(\frac{1}{qd}\right)
\end{equation}
This term has sign opposite to the one of  a crystalline
membrane, \ref{delta}, and is much smaller than the latter. 

In presence of strongly interacting long-wavelength
fluctuations, scaling considerations are extremely useful\cite{Ma}. Let us assume that the behavior of the renormalized bending rigidity $\kappa_R(q)$ at small $q$ is determined by
some exponent $\eta$, $\kappa_R(q) \propto q^{-\eta}$ yielding
\begin{equation}
G(q)=\frac{A}{q^{4-\eta}q_0^\eta},~~~~~~<|\vec{n_{\vec{q}}}|^2>=\frac{A}{q^{2-\eta}q_0^\eta}
\label{Gq}
\end{equation}
where the parameter $q_0=\sqrt{Y/\kappa}$
of the order of $d^{-1}$ is introduced to make $A$ dimensionless. One can assume also a
renormalization of the effective Lam\'e coefficients $\lambda_R(q)$, $\mu_R(q)$ $ \propto q^{\eta_u}$
which means
\begin{equation}
<u_{\alpha \vec{q}}^*u_{\beta \vec{q}}^{ }>\propto \frac{1}{q^{2+\eta_u}}
\label{etau}
\end{equation}
Finally, we assume that anharmonicities change \ref{h2} into 
\begin{equation}
<h^2>\propto L^{2\zeta}
\end{equation}
The values $\eta$,$\eta_u$ and $\zeta$ are similar to critical exponents in the theory of critical
phenomena. They are not independent\cite{nelsonbook,bookKats}
\begin{equation}
\zeta=1-\eta/2 ,~~~~~\eta_u=2-2\eta
\end{equation}
The exponent $\eta_u$ is positive if $0<\eta<1$.
The so-called Self-Consistent-Screening-Approximation\cite{ledoussal} gives $\eta \approx 0.82$ whereas a more accurate renormalization group approach\cite{mohanna} yields $\eta \approx 0.85$.
This means that, interactions make out-of-plane
phonons harder  and in-plane phonons  softer.

The temperature dependence of the constant $A$ in \ref{G} can be found
from the assumption that  \ref{G0} and \ref{Gq} should match at $q=q^*$, giving
$A=\alpha\left(T/\kappa\right)^\zeta$
where $\alpha$ is a dimensionless factor of the order of one. 

Now we are ready to discuss the possibility of long-range crystal order in 2D systems at finite temperatures.
The true manifestation of long-range order is the existence of delta-function
(Bragg) peaks in diffraction experiments. The scattering intensity is proportional to the structure factor
\begin{equation}
S(\vec{q})=\sum_{nn'}\sum_{jj'} \left<\exp\left[ i \vec{q} \left(\vec{R}_{nj}- \vec{R}_{n'j'} \right) \right] \right>
\label{S}
\end{equation}
that can be rewritten as
\begin{equation}
S(\vec{q})=\sum_{nn'}\exp\left[ i \vec{q} \left(\vec{r}_{n} - \vec{r}_{n'}\right)\right] \sum_{jj'} \exp\left[ i \vec{q} \left(\vec{\rho}_{j}- \vec{\rho}_{j'} \right) \right] 
W_{nj,n'j'}
\label{S1}
\end{equation}
where
\begin{equation}
 W_{nj,n'j'}=\left<\exp \left[ i \vec{q} \left(\vec{u}_{nj} - \vec{u}_{n'j'}\right)\right]\right>
\end{equation}
In  3D crystals, one can assume that the
displacements $\vec{u}_{nj}$ and $\vec{u}_{n'j'}$ are not correlated for $|\vec{r}_n-\vec{r}_{n'}|\rightarrow\infty$ so that
\begin{equation}
 W_{nj,n'j'}=\left< \exp \left( i \vec{q} \vec{u}_{nj} \right) \right> \left< \exp \left(- i \vec{q} \vec{u}_{n'j'} \right) \right>\equiv m_j(\vec{q})m^*_{j'}(\vec{q})
\label{W}
\end{equation}
where $m_j(\vec{q})$  are Debye-Waller factors that are independent
of $n$ due to translational invariance. Therefore, for $\vec{q}=\vec{g}$  (reciprocal lattice
vectors), where $\exp{ (i\vec{q}\vec{r}_n)}=1$, the contribution to $S(\vec{q})$ is proportional to $N^2_0$,
whereas for a generic $\vec{q}$ it is of the order of $N_0$. The Bragg peaks at $\vec{q}=\vec{g}$ are,
therefore, sharp; thermal fluctuations decrease their intensity (by the Debye-Waller factor) but do not broaden the peaks. The observation of such 
peaks is an experimental manifestation of long-range crystal order. In 2D the correlation functions of atomic
displacements do not vanish as $|\vec{r}_n-\vec{r}_{n'} |\rightarrow \infty$. Indeed, in the continuum limit,
$\vec{u}_{nj}\rightarrow (\vec{u}(\vec{r}),h(\vec{r}))$  and we have
\begin{equation}
\left< \left[ h(\vec{r})-h(\vec{r}')\right]^2\right> = 2 \sum_{\vec{q}} \left< |h(\vec{q}|^2\right> \left[ 1-\cos \left(\vec{q} \left( \vec{r}- \vec{r}'\right) \right) \right] \sim |\vec{r}- \vec{r}'|^{2\zeta}
\end{equation}
\begin{equation}
\left<\left[\vec{u}(\vec{r})-\vec{u}(\vec{r'})\right]^2\right>=2 \sum_{\vec{q}} \left<|\vec{u}(\vec{q}|^2\right> \left[1-\cos\left(\vec{q}\left( \vec{r}- \vec{r}'\right) \right) \right] \sim |\vec{r}- \vec{r}'|^{\eta_u}
\end{equation}
after substitutions of \ref{Gq},\ref{etau}\cite{bookKats}.
Thus the approximation \ref{W} does not apply.
As a result, the sum over
$n'$ at a given $n$ is convergent, and $S(\vec{q}=\vec{g})\propto N_0$; instead of a delta-function
Bragg peak we have a sharp maximum
of finite width. This means that,
rigorously speaking, the statement that 2D crystals cannot exist at finite temperatures
 is correct. However, the structure factor of graphene still has sharp maxima at $\vec{q}=\vec{g}$ and the crystal
lattice can be determined from the positions of these maxima. 
In this restricted
sense, 2D crystals do exist, and graphene is a prototype example
of them.

It was found experimentally by transmission electron microscopy, that freely suspended graphene at room temperature is rippled\cite{Meyer}. The existence of these  thermally induced ripples  motivated our 
atomistic Monte Carlo simulations\cite{our,wave} summarized in the next section.

\section{Atomistic simulations of structural and thermal properties of graphene}

As discussed before the thermal properties of 2D crystals are determined by longwavelength fluctuations. Therefore, one needs to deal with large enough systems to probe the interesting regime of strongly interacting fluctuations. This requirement rules out, in practice, first principle approaches in favor of accurate empirical potentials.
The unusual structural aspects of graphene, 
 make it desirable to describe different structural and bonding  configurations, beyond the harmonic approximation, by means of a unique interatomic potential.  Bond order potentials are a class of empirical interatomic potentials  designed for this purpose (see \cite{Los2005} and references therein). They aim at describing also anharmonic effects and the possible breaking and formation of bonds in structural phase transitions. They allow to study without further adjustment of parameters,  all carbon structures, including the effect of defects, edges and other structural changes, also as a function of temperature as well as phonon spectra. We have used the so-called  long-range carbon bond order potential LCBOPII\cite{Los2005}. Its main innovative feature is the treatment of interplanar van der Waals interactions, that allows to deal with graphitic structures. To calculate equilibrium properties as a function of temperature, we have performed Monte Carlo simulations either at constant volume or constant pressure. We first discuss the results for correlations functions that can be directly compared to the scaling behavior discussed previously. Then, we report the temperature dependence of several structural properties of graphene.  Lastly we discuss the melting of graphene in relation to its 3D counterpart, graphite, and to 2D models of melting.  

\subsection{Structural properties and scaling}

\begin{figure}
\includegraphics[width=0.5\linewidth,angle=0,clip]{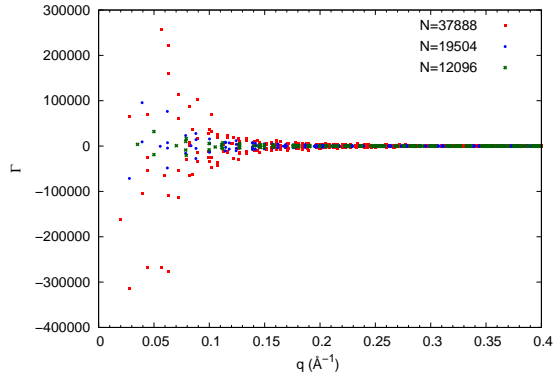}
\caption{The function $\Gamma(q)$ for three values of $N$}
\label{Fig:Gamma}
\end{figure}

We compare the results of atomistic Monte Carlo simulations to the scaling behavior of $G(q)$ (\ref{Gq}). From \ref{Gq}, one can see that $G(q)$ can be calculated in  two ways, either by calculating directly the correlator of $h(\vec{r})$ or of the normal $\vec{n}(\vec{r})$. In doing this, it is important to have $h(\vec{r})$ and $\vec{n}(\vec{r})$ calculated at lattice sites smoothened by averaging over the neighbors, as described in detail in Ref.\cite{Zak2010}. Only by such a procedure one verifies numerically \ref{nn} for$q<10$ nm$^{-1}$ which gives the limit of applicability of a continuum description to graphene. The interesting regime is $q<<q^*$ (\ref{q*}). For graphene at room temperature $q^*=2.4$ nm$^{-1}$. Since simulations are done for samples of dimension $L_x \times L_y$ with periodic boundary conditions, the smallest values of $q$ that can be reached are $2\pi/L_x$ and $2\pi/L_y$. For the largest samples, we have found that straightforward Monte Carlo simulations based on individual atomic moves, could not provide enough sampling for the smallest wavevectors.
For this reason in our first paper on ripples in graphene\cite{our} we were not able to check the scaling laws in the anharmonic regime. Later, we have reached this regime by devising a numerical technique that we have called wave moves, where collective sinusoidal long wavelengths displacements of all atoms where added in the Monte Carlo equilibration procedure\cite{wave}. 
% figure: /tex/papers/zakharchenko/exp_model/GqLCBOPII.agr with xaxis up to 4.
\begin{figure}
\includegraphics[width=0.5\linewidth,angle=0,clip]{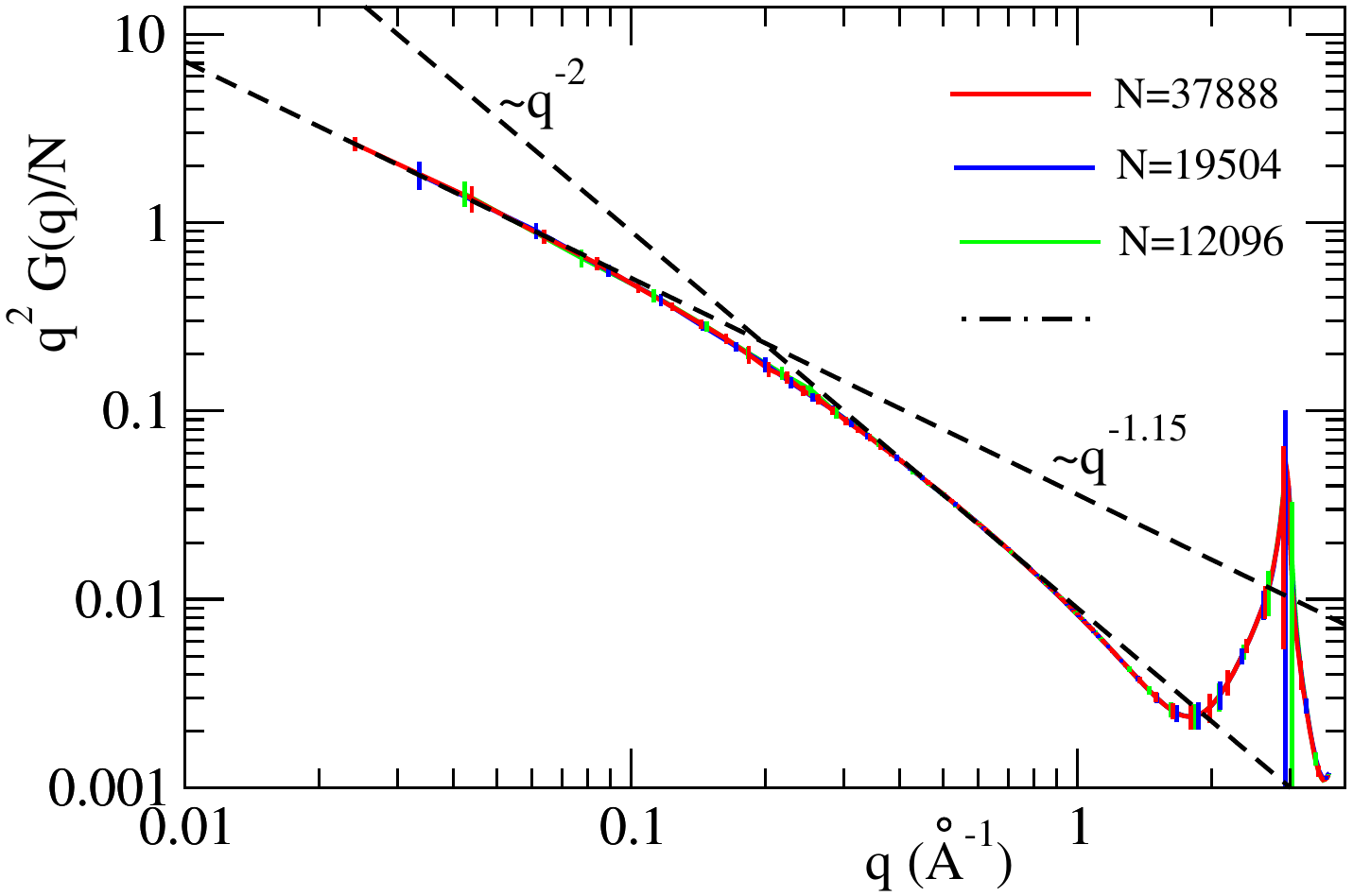}
\caption{Normal-normal correlation function $q^2 G(q)$ for three samples with indicated number of atoms $N$. For the largest, $N=37888$, $L_x=314.82$ \AA, $L_y=315.24$ \AA. Adapted from Ref.\cite{wave}}
\label{Fig:GQ}
\end{figure}
In \ref{Fig:GQ} we present $<|\vec{n}_{\vec{q}}|^2>=q^2 G(q)$ calculated with wave moves  which displays a clear change of the slope $\ln (G(q))$ versus $\ln (q)$ around $q\sim q^*$. Notice that  $<|\vec{n}_{\vec{q}}|^2>$ grows for  $q> 10$ nm$^{-1}$ reaching a maximum at the first Bragg peak. According to the phenomenological theory described before, the change of scaling behaviour at $q<<q^*$  is related to the coupling of in-plane and out-of plane fluctuations. To check this, we also show in \ref{Fig:Gamma} the correlation function $\Gamma(q)$
\begin{equation}
\Gamma(q)=<(u_x)_{\vec{q}}(h^2)_{-\vec{q}}>
\end {equation}
which becomes almost zero at $q > q^*$. For smaller samples the coupling is reduced as expected for a property that is determined by the region of long wavelengths fluctuations. 

\begin{figure}
\includegraphics[width=0.5\linewidth,angle=0,clip]{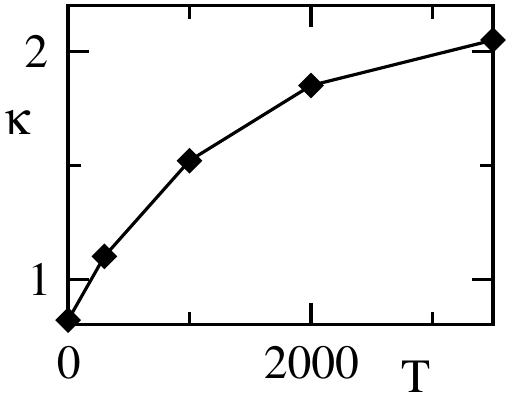}
\caption{Temperature dependence of $\kappa$ as found by fitting  $<|n_{\vec{q}}|^2>$  to \ref{Gq}. Adapted from Ref.\cite{our}}
\label{Fig:KT}
\end{figure}

The temperature dependence of the bending rigidity $\kappa(T)$ can be extracted from the $<|n_{\vec{q}}|^2>$ using \ref{nnT}. The results\cite{our} are shown in \ref{Fig:KT} where one can see the rapid growth with temperature. This effect should not be confused with the correction $\delta\kappa$ of \ref{delta} since the latter is strongly $q$-dependent. The temperature dependence of $\kappa$ of \ref{Fig:KT} cannot be described within the Self-Consistent Screening Approximation for our model Hamiltonian \ref{eq:Free_F_1}\cite{roldan}. 

The temperature dependence of $\kappa$, as that of all parameters of phonon spectra\cite{Cowley}, is an anharmonic effect that goes beyond the 
model \ref{eq:Free_F_1}, namely it does not result from the coupling of acoustic out-of plane phonons with acoustic in-plane phonons only. Other anharmonicities, like coupling to other phonons have to be invoked. This is an example of effects  that can be studied within atomistic simulations but not within the elasticity theory. The potential energy given by LCBOPII includes by construction anharmonic effects.

\begin{figure}
\caption{Temperature dependence of the lattice parameter $a$ calculated by Monte Carlo simulations at zero pressure. Adapted from Ref.\cite{prlkostya}}
\includegraphics[width=0.5\linewidth,angle=0,clip]{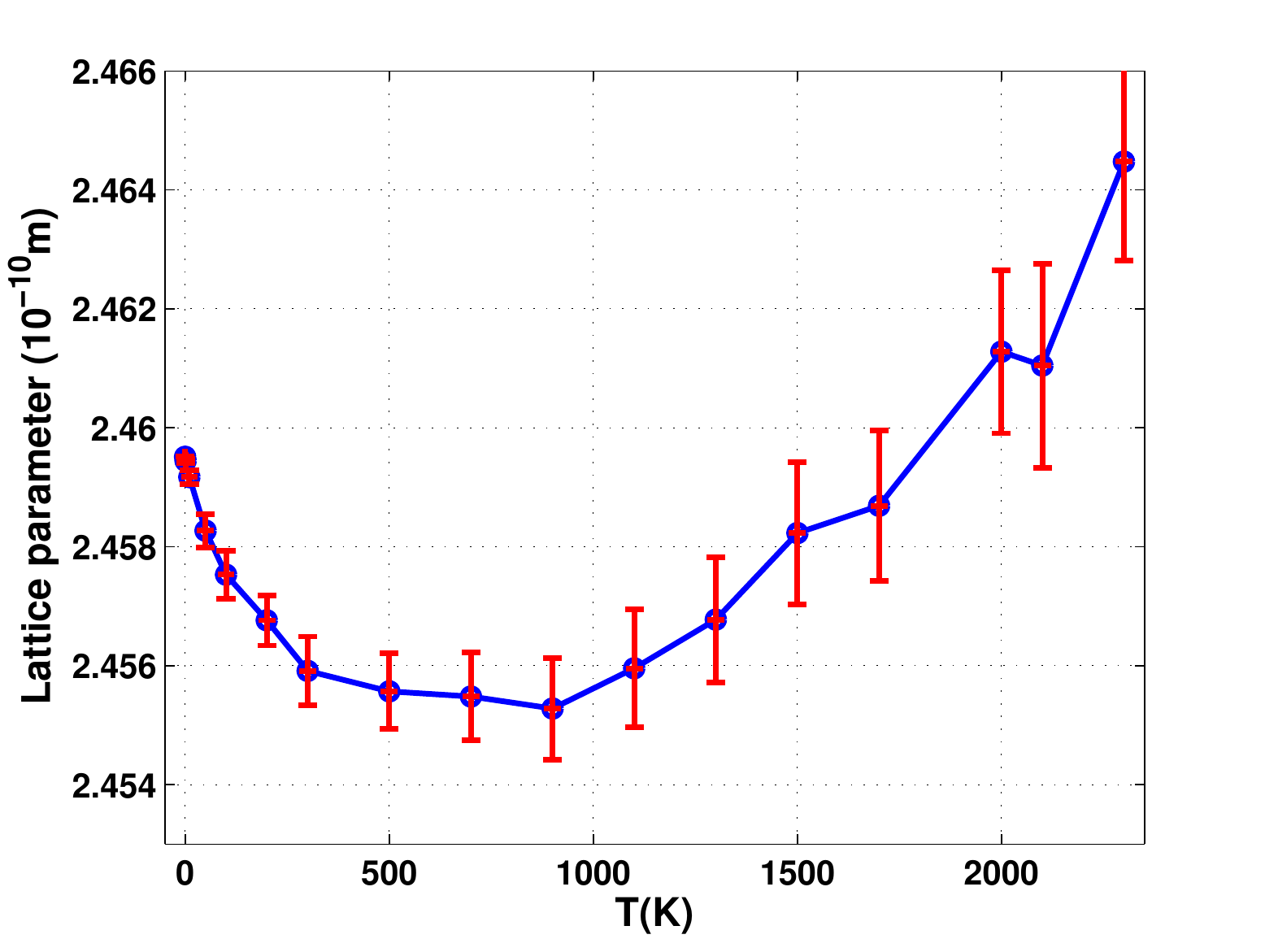}
\label{Fig:a_T}
\end{figure}

\begin{figure}
\includegraphics[width=0.5\linewidth,angle=0,clip]{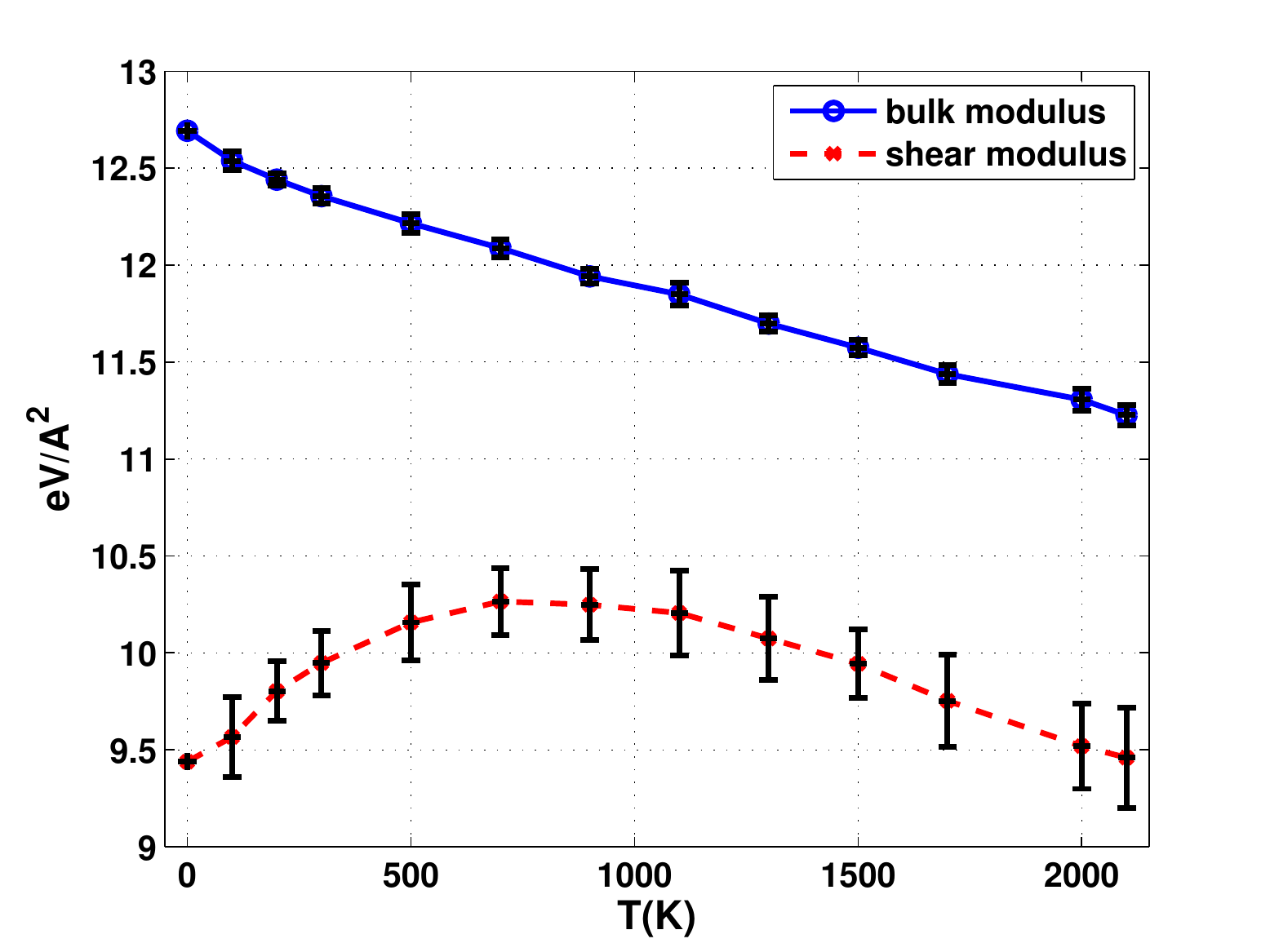}
\caption{Temperature dependence of the bulk modulus $b_A$ and shear modulus $\mu$. Adapted from Ref.\cite{prlkostya}}
\label{Fig:bA-mu_T}
\end{figure}

 Purely anharmonic effects are the temperature dependence of lattice parameter and elastic moduli. The temperature dependence\cite{prlkostya} of the lattice parameter $a$ is shown in \ref{Fig:a_T} and those of the shear modulus $\mu$ and adiabatic bulk modulus $b_A=\lambda+\mu$ in \ref{Fig:bA-mu_T}. The most noticeable  feature in \ref{Fig:a_T} is the change of sign of $da/dT$, namely a change from thermal contraction to thermal expansion around 1000 K.  

\begin{figure}
\includegraphics[angle=0,scale=0.5]{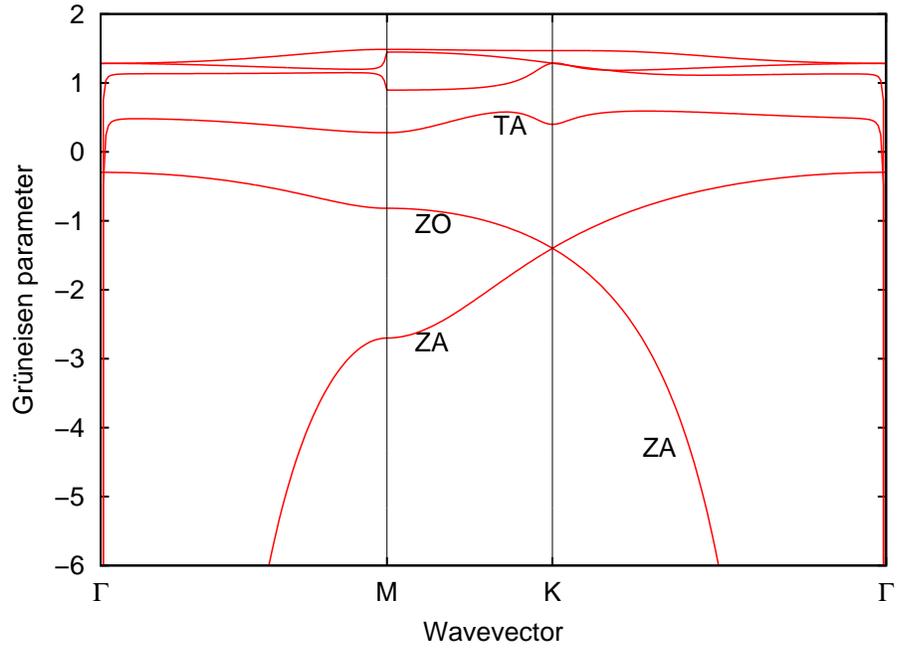}
\includegraphics[angle=0,scale=0.5]{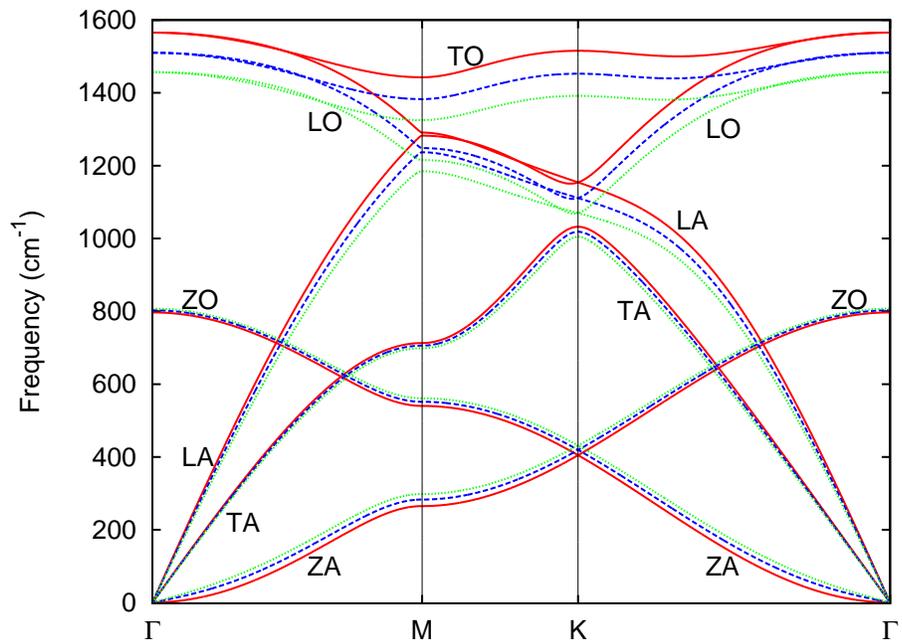}
\caption{Top) Gr\"uneisen parameters calculated for graphene with LCBOPII\cite{Karssemeijer}; bottom) Phonon spectrum of graphene for the equilibrium value of the interatomic distance 1.42 \AA~ (red solid), and two larger values, $1.43$ \AA~ (blue dashed) and  $1.44$ \AA~ (green dotted). Courtesy of L.J. Karssemeijer}
\label{Fig:Grungraphene}
\end{figure}

Usually thermal expansion is described in the quasiharmonic 
approximation\cite{Cowley} where the free energy is written as in the harmonic approximation but with volume dependent phonon frequencies $\omega_\lambda$. This dependence is described by the Gr\"uneisen parameters 
\begin{equation}
\gamma_\lambda=-\frac{\partial \ln \omega_{\lambda}}{\partial \ln \Omega}
\end{equation}
where $\Omega$ is the volume (the area for 2D systems). In most solids, phonon frequencies grow under compression, which corresponds to positive Gr\"uneisen parameters and thermal expansion. Graphene and graphite are however exceptional, as illustrated in \ref{Fig:Grungraphene} presenting the corresponding calculation with LCBOPII\cite{Karssemeijer}. One can see that both the $ZA$ and
$ZO$ branches have $\gamma<0$ almost in the whole Brillouin zone as found already, within density functional calculations\cite{MounetMarzari}. 

Experimentally, graphite has a negative thermal expansion coefficient up
to 700K \cite{Steward}. This behavior has been
explained in the quasiharmonic approximation in Ref.\cite{MounetMarzari}. For graphene, they  predicted negative $da/dT$ at all temperatures. Negative thermal expansion of graphene
at room temperature has been confirmed experimentally\cite{Bao}.
The linear thermal expansion coefficient
was about -10$^{-5}$ K$^{-1}$, a very large negative value. According to
the quasiharmonic theory, it was found to be
more or less constant up to temperatures of the order of at least 2000 K ,in contrast to the atomistic simulations of \ref{Fig:a_T}. Thus the change of sign of $da/dT$  should be attributed to self-anharmonic effects\cite{Cowley}, namely to direct effects of phonon-phonon interactions. 
Very recently, it was confirmed experimentally
that $da/dT$, while remaining negative, decreases in modulus with increasing
temperature up to 400K\cite{Yoon}, which can be considered as a partial confirmation of our prediction. 

Also the temperature dependence of the shear modulus $\mu$, shown in \ref{Fig:bA-mu_T} is anomalous since typically $d\mu/dT<0$ at any temperature. The change of sign of $d\mu/dT<0$ occurs roughly at the same temperature of $da/dT$. The room-temperature values
of the elastic constants are  $\mu \approx 10$ eV/\AA$^2$ and $b_A \approx 12$ eV/\AA$^2$. 
The corresponding Young modulus $Y$ 
lies within the error bars of the experimental
value\cite{Science} $Y\approx 340 \pm50$ Nm$^{-1}$. The Poisson ratio $\nu=(b_A-\mu)/(b_A+\mu)$ is found to be very small, of the order of $0.1$.

\subsection{Melting of graphene}

Melting in 2D is usually described in terms of creation of topological defects, like unbound disclinations that destroy orientational order and unbound dislocations that destroy translational order\cite{4}. In the hexagonal lattice of graphene, typical disclinations are pentagons (5) and heptagons (7) while dislocations are 5-7 pairs. Our atomistic simulations\cite{ourmelting} have given an unexpected scenario of the melting of graphene as the decomposition of the 2D crystal in a 3D network of 1D chains. A crucial role in the melting process is played by the Stone-Wales (SW) defects, non-topological defects with a 5-7-7-5 configuration. The SW defects have the smallest formation energy and start appearing spontaneously at about 4200 K. It is the clustering of SW defects that triggers the spontaneous melting around 4900 K in our simulations.
\begin{figure}
\includegraphics[width=0.5\linewidth,angle=0,clip]{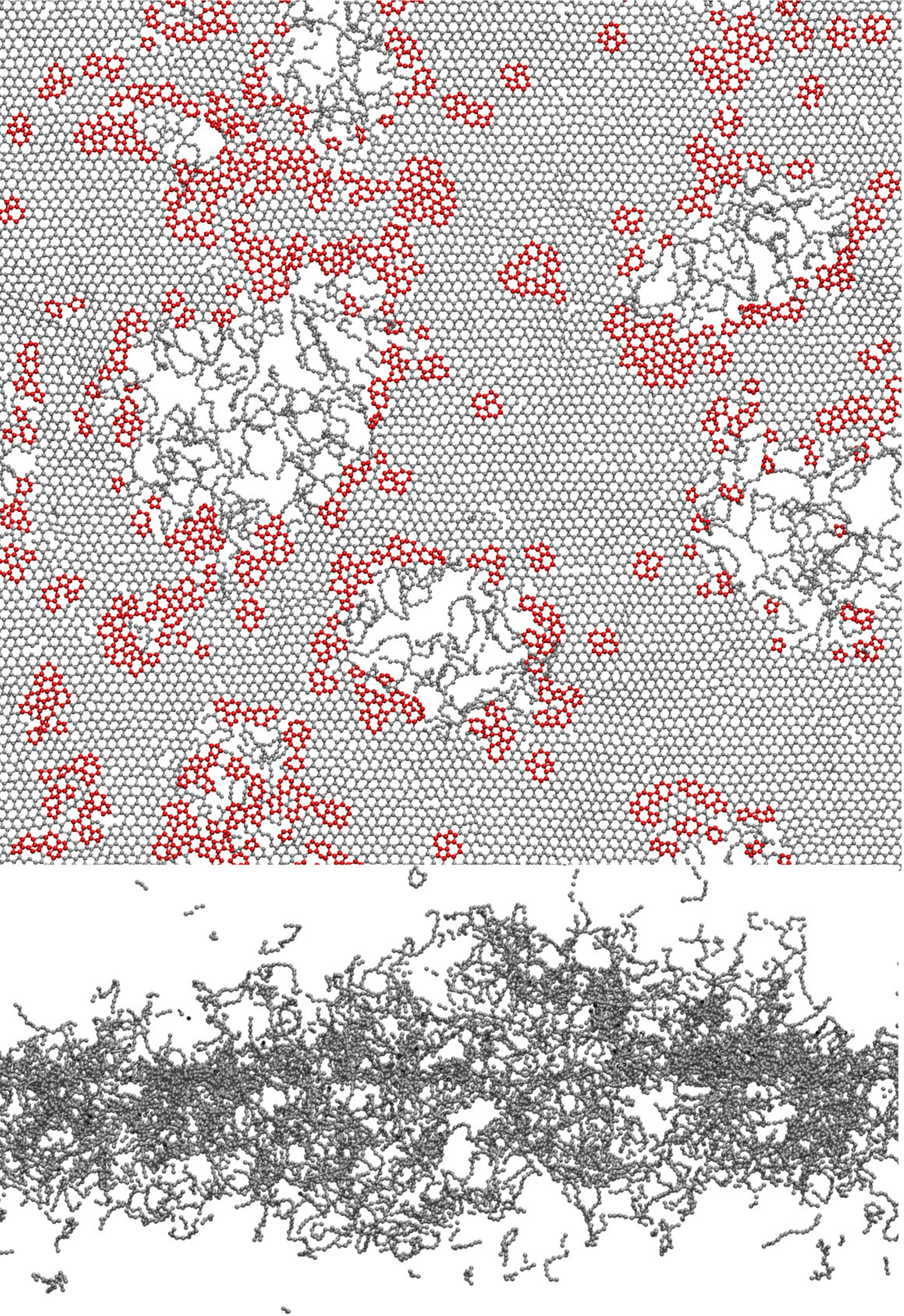}
\caption{Structure of graphene in the first phase of melting (top) and when molten (bottom) at $T=5000$ K. Adapted from Ref.\cite{ourmelting}}
\label{Fig:meltingstructure}
\end{figure}

In \ref{Fig:meltingstructure} we show a typical configuration of graphene on the way to melting at 5000 K. The coexistence of crystalline and molten regions indicates a first order phase transition. The most noticeable features are the puddles of graphene that have molten into chains. The molten areas are surrounded by disordered 5-7 clusters, resulting from the clustering and distortion of SW defects. Isolated and pairs of SW defects are also present whereas we never observe isolated pentagons, heptagons or 5-7 dislocations. Contrary to graphite where melting is initiated by interplanar covalent bond formation, in graphene it seems that 5-7 clusters act as nuclei for the melting. By close inspection, we find that regions with 5-7 clusters favor the transformation of three hexagons into two pentagons and one octagon that we never see occurring in the regular hexagonal lattice far from the 5-7 clusters. The large bonding angle in octagons, in turn, leads to  the proliferation of larger rings. Due to the weakening of the bonds with small angles in the pentagons around them,  these larger rings tend to detach from the lattice and form chains. 

When melting is completed the carbon chains form an entangled 3D network with a substantial amount of three-fold coordinated atoms, linking the chains.  The molten phase is similar to the one found for fullerenes and nanotubes(see \cite{ourmelting} and references therein). Therefore, the structure of the high temperature phase reminds rather a polymer gel than a simple liquid, a quite amazing fact for an elemental substance. 

The closest system to graphene is graphite. The melting temperature $T_m$ of graphite has been extensively studied experimentally at pressures around $10$ GPa and the results present a large spread between $4000$ K and $5000$ K \cite{13}. With LCBOPII, free energy calculations give $T_m = 4250$ K, almost independent of pressure between $1$ and $20$ GPa (see references in \cite{ourmelting}. At zero pressure, however,  graphite sublimates before melting at $3000$ K \cite{13}. Monte Carlo simulations with LCBOPII at zero pressure show that, at $3000$ K, graphite sublimates through detachment of the graphene layers. 
The melting of graphene in vacuum that we have studied here can be thought of as the last step in the thermal decomposition of graphite. Interestingly, formation of carbon chains has been observed in the melt zone of graphite under laser irradiation\cite{Hu}.
Although the temperature $T=4900$~K of spontaneous melting represents an upper limit for $T_m$, our simulations suggest that $T_m$ of graphene at zero pressure is higher than that of graphite. 

\section{Conclusions}
We have shown by comparing the results of atomistic simulations to the theory of membranes based on a continuum approach, that graphene can indeed be considered a prototype of 2D membrane and that atomistic studies can be used to evaluate accurately the scaling properties, including scaling exponents and cross-over behavior. Conversely, the melting of graphene is determined rather by the peculiarities of the carbon-carbon bond and the high stability of carbon chains than as a generic model for melting in 2D. 

\acknowledgements
We are grateful to our collaborators, Jan Los, Kostya Zakharchenko and Lendertjan Karssemeijer.
This work is supported by FOM-NWO, the Netherlands.

\end{document}